\documentclass{article}

\usepackage{microtype}
\usepackage{graphicx}
\usepackage{subfigure}
\usepackage{booktabs} % for professional tables
\usepackage[inline,shortlabels]{enumitem}
\usepackage{hyperref}

\usepackage[accepted]{tom2023}

\usepackage{amsmath}
\usepackage{amssymb}
\usepackage{mathtools}
\usepackage{amsthm}

\usepackage[capitalize,noabbrev]{cleveref}

\theoremstyle{plain}

\theoremstyle{definition}

\theoremstyle{remark}

\usepackage[textsize=tiny]{todonotes}

\icmltitlerunning{Evaluating Summary Statistics with Mutual Information for Cosmological Inference}

\begin{document}

\twocolumn[
\icmltitle{Evaluating Summary Statistics with Mutual Information for Cosmological Inference}



\icmlsetsymbol{equal}{*}

\begin{icmlauthorlist}
\icmlauthor{Ce Sui}{yyy}
\icmlauthor{Xiaosheng Zhao}{yyy}
\icmlauthor{Tao Jing}{yyy}
\icmlauthor{Yi Mao}{yyy}
\end{icmlauthorlist}

\icmlaffiliation{yyy}{Department of Astronomy, Tsinghua University, Beijing 100084, China}

\icmlcorrespondingauthor{Ce Sui}{suic20@mails.tsinghua.edu.cn}
% \icmlcorrespondingauthor{Yi Mao}{ymao@tsinghua.edu.cn}

\icmlkeywords{Statistical Inference, Mutual information}

\vskip 0.3in
]



\printAffiliationsAndNotice{}  % leave blank if no need to mention equal contribution

\begin{abstract}
The ability to compress observational data and accurately estimate physical parameters relies heavily on informative summary statistics. In this paper, we introduce the use of mutual information (MI) as a means of evaluating the quality of summary statistics in inference tasks. MI can assess the sufficiency of summaries, and provide a quantitative basis for comparison. We propose to estimate MI using the Barber-Agakov lower bound and normalizing flow based variational distributions. To demonstrate the effectiveness of our method, we compare three different summary statistics (namely the power spectrum, bispectrum, and scattering transform) in the context of inferring reionization parameters from mock images of 21~cm observations with Square Kilometre Array. We find that this approach is able to correctly assess the informativeness of different summary statistics and allows us to select the optimal set of statistics for inference tasks.
% The ability to compress observational data and accurately estimate physical parameters relies heavily on informative summary statistics. In this paper, we introduce the use of mutual information (MI) as a means of evaluating the quality of summary statistics in inference tasks. MI can assess the sufficiency of summaries, and provide a quantitative basis for comparison. We propose to estimate MI using the Barber-Agakov lower bound and normalizing flow based variational distributions. To demonstrate the effectiveness of our method, we compare three different summary statistics (namely the power spectrum, bispectrum, and scattering transform) in the context of inferring reionization parameters from mock SKA images. We find that this approach is able to correctly assess the informativeness of different summary statistics and allows us to select the optimal statistic for our inference task.
\end{abstract}

\section{Introduction}
\label{sec:intro}
Statistical inferences in cosmology consist of two parts. Firstly, summary statistics are selected to extract relevant information from raw observed data. These statistics are then used to infer parameters for a given physical model. The choice of summary statistics is crucial for obtaining better constraints on these physical parameters. Many summary statistics have been proposed to extract information from various types of astronomical datasets. For Cosmic Microwave Background (CMB) studies, analyses have largely focused on the power spectrum \citep{ade2016planck}. However, other cosmological fields, such as the 21 cm signal, are expected to be highly non-Gaussian, necessitating more informative summary statistics. As a result, new summary statistics have been proposed in 21 cm cosmology, such as the bispectrum \citep{Bispectrum,Shimabukuro_bs_eor_infer} and Minkowski Functionals \citep{MFs,MFs2}. Furthermore, neural networks are being explored for learning summaries from input images in the context of cosmological inference \citep{xs-cnn-delfi,21cm-RNN,imnn}.

% An important task is therefore predicting the constraining power of different statistics on target physical parameters and make comparisons. There is no universal way to compare different summary statistics in a given inference task, as summary statistics are often derived from different frameworks. Traditional comparisons of new statistics often use posterior comparisons and Fisher analysis \citep{Watkinson_bs_eor_infer,Shimabukuro_bs_eor_infer_fisher,xs-cnn-delfi}, both of which typically consider only one set of fiducial parameters. The former one is based on inference results on a mock observation and examines the posteriors of different statistics to evaluate their effectiveness. The latter one requires a tractable likelihood and calculate the fisher information at the fiducial parameter. Predictions from these two methods consider one point at parameter space and may not reflect the overall performance of the statistics. One way to evaluate the statistics in large parameters space is through regression performance, i.e. by checking the optimal performance that the statistics can achieve in predicting parameters in the whole parameter ranges \citep{xs-cnn-delfi,21cm-RNN}. However, this method only considers point estimates.

An important task is to predict and compare the effectiveness of different summary statistics in constraining target physical parameters. There is no universal way to do that in a given inference task, as summary statistics are often derived from different frameworks. Traditional approaches for comparing new statistics often involve posterior comparisons and Fisher analysis \citep{Watkinson_bs_eor_infer,Shimabukuro_bs_eor_infer_fisher,xs-cnn-delfi}. These methods typically consider one set of fiducial parameters. The former approach relies on inference results from mock observations and assesses the effectiveness of different statistics by examining their posteriors. The latter approach often uses Gaussian assumptions and calculates the Fisher information at the fiducial parameter. However, predictions obtained from these methods only consider a single point in the parameter space, which might not fully capture the overall performance of the statistics. One way to evaluate statistics across a large parameter space is through regression performance. This involves assessing the optimal performance that the statistics can achieve in predicting parameters across the entire parameter range \citep{xs-cnn-delfi,21cm-RNN}. However, this approach only considers point estimates and does not provide a comprehensive analysis of statistical performance.

A similar problem is extensively studied in the machine learning community in the context of representation learning. The goal of representation learning is to find a low-dimensional representation that preserves most of the information from the original data. It has been shown that such task can be framed as a problem of maximizing the mutual information (MI) between the learned summaries and the original data \citep{infomax,infonce}. This suggests that we may use a similar metric to quantify the effectiveness of different summary statistics.

% In this work, we propose to compare different summary statistics by estimating the mutual information between summary statistics and target parameters in a given inference task. This method provides a quantitative comparison between different statistics and evaluates their effectiveness by considering the statistical dependence of mock observation and parameters. It considers the total parameter space and can help us can gain a better understanding of how summary statistics capture relevant information for inference.

In this study, we introduce a novel approach to compare different summary statistics by estimating the mutual information between the statistics and target parameters in a given inference task. This method enables a quantitative comparison of the effectiveness of various statistics by considering their statistical dependence on parameters. Unlike many previous methods that focus on a single point in the parameter space, our approach takes into account the entire parameter range, providing a comprehensive evaluation of how well the summary statistics capture the necessary information for inference. 

\section{Method}
\label{sec:method}
In statistical inference we seek to estimate the parameters $\theta$ of a physical model given some input observation $x$. In Bayesian inference, this requires estimating the posterior distribution $p(\theta|x)$. However, the original observation is typically high-dimensional and contains a substantial amount of irrelevant information. To address this, we often utilize summary statistics to compress the data into more compact representations $s$ that contain the most critical information about the parameter.

To achieve optimal performance, summary statistics should capture all the relevant information contained in the original observation $x$ regarding the parameter of interest. This leads to the definition of sufficient statistics as those that satisfy the condition
\begin{equation}
p(\theta|x,s)=p(\theta|s),
\label{eq:suf_sta}
\end{equation}
which implies that once the summary statistics are given, the original data does not provide additional information. If a summary statistic closely resembles the sufficient statistic, it is likely to provide more accurate and reliable information about the underlying parameter.

\subsection{Mutual information as a probe of Sufficient Statistics
}

Mutual information is a fundamental concept in statistical inference and information theory that measures the amount of information one random variable contains about another. Mutual information is defined as
\begin{equation}
\begin{aligned}
\mathrm{I}({x}; {y})&=KL(p({x}, {y})||p({x}) p({y})) \\
&=\mathbb{E}_{p({x}, {y})}\left[\log \frac{p({x}, {y})}{p({x}) p({y})}\right],
\label{eq:MI}
\end{aligned}
\end{equation}
where ${x}$ and ${y}$ are random variables, and $p({x}, {y})$, $p({x})$, and $p({y})$ are their joint and marginal probability distributions, respectively. Mutual information quantifies the difference between the joint distribution and the product of the marginal distributions using the Kullback-Leibler divergence (KLD). If $x$ and $y$ are independent, their mutual information is zero, indicating that knowing $x$ does not provide any information about $y$. Conversely, if $x$ contains information about $y$, mutual information is non-zero, and a higher value indicates a stronger dependence between the two variables.

Mutual information can also be used to define sufficient statistics. For Bayesian inference, we can show that Equation~\ref{eq:suf_sta} is equivalent to
\begin{equation}
I(\theta;x)=I(\theta;s(x)),
\label{eq:suf_sta_mi}
\end{equation}
which implies that sufficient statistics $s$ contain all the information about $\theta$ in the original observation $x$. Thus, by measuring the mutual information between a summary statistic and the target parameters, we can evaluate the sufficiency of the statistic. This interpretation provides a powerful tool for selecting summary statistics and assessing their suitability for use in statistical inference.

\subsection{Mutual information estimation
}

Estimating mutual information is a challenging task in practice as it requires the computation of the KLD between complex distributions, which are often unknown. For many cosmological inference tasks we do not have access to a tractable joint distribution of the physical parameters and the data summaries, $p(\theta,x)$. Hence, it is infeasible to directly evaluate the mutual information between them. However, we can use variational distributions to approximate the real distributions, and obtain variational bounds of mutual information.

Assuming that we have a variational distribution $q(\theta|s)$, we can utilize it to replace the actual conditional distribution $p(\theta|s)$ and prove that it generates a lower bound on MI. Specifically, we have
\begin{equation}
\begin{aligned}
I(\theta ;s )= & \mathbb{E}_{p({\theta} , {s})}\left[\log \frac{p({\theta} | s)}{p({\theta})}\right] \\
=& \mathbb{E}_{p({\theta} , {s})}\left[\log \frac{q(\theta | s)p({\theta} | {s})}{p({\theta})q(\theta | s)}\right] \\
= & \mathbb{E}_{p(\theta, s)}\left[\log \frac{q(\theta | s)}{p(\theta)}\right] \\
&+\mathbb{E}_{p(s)}[K L(p(\theta | s) \| q(\theta | s))] \\
\geq & \mathbb{E}_{p(\theta, s)}[\log q(\theta | s)]+h(\theta),
\end{aligned}
\label{eq:ba-bound}
\end{equation}

where $h(\theta)=-\mathbb{E}_{p(\theta)}[\log p(\theta)]$ is the differential entropy of $\theta$ and the last inequality is due to the non-negativity of the KLD. This is often referred as the Barber-Agakov lower bound \citep{ba-bound,VBofMI}. It is evident from this equation that the replacement of the real conditional distribution with a variational model results in a lower bound on MI, which is only tight when $p(\theta | s)=q(\theta | s)$. In cosmological inference problems, calculating the differential entropy of the prior distribution of $\theta$ is relatively straightforward as it is typically tractable. For the first component, we can fit a highly flexible generative model to a large number of parameter-statistic pairs, yielding a variational distribution that approximates the actual distribution closely. In this work, we use a masked autoregressive flow (MAF; \citealp{MAF}) as the variational distribution. We optimize MAFs by minimizing the negative log-probability, which is equivalent to maximizing the lower bound. In principle, alternative loss functions \citep[e.g.][]{score_match_maf} can also be employed.

\subsection{Relation with other methods}
\label{subsec:unified_MI}
We can also consider other methods commonly used for comparing data summaries as processes of MI estimation. For instance, we can write MI as
\begin{equation}
\begin{aligned}
I(\theta ;s)=&\mathbb{E}_{p(s)}[K L(p(\theta | s) \| p(\theta))]. \\
\end{aligned}
\label{eq:MI-2}
\end{equation}
 Using this expression, we can estimate MI by assuming $p(\theta | s) = \hat{p}(\theta | s_0)$, where we replace the true conditional distribution with a posterior distribution that we estimate at a particular observation $s_0$. In this case, the estimated MI becomes $\hat{I}(\theta ;s)=K L(\hat{p}(\theta | s_0) \| p(\theta))$, which measures the difference between the estimated posterior and the prior. This is similar to comparing different statistics by examining their posteriors on one mock observation. However, this estimation has high variance since it uses only one estimated posterior distribution to represent the conditional distribution.

Regression performance can also be utilized to formulate a Barber-Agakov lower bound. In one dimensional case if we let $q(\theta | s) = \mathcal{N}\left(f(s), \mathbb{E}_{p(\theta, s)}\left[(\theta-f(s))^2\right]\right)$ in equation~\ref{eq:ba-bound}, where $f(s)$ is a estimator trained to predict $\theta$ from given summaries statistics $s$, then:
\begin{equation}
\begin{aligned}
I(\theta ;s) \geq &  \mathbb{E}_{p(\theta, s)}[\log q(\theta | s)]+h(\theta) \\
= & h(\theta)-1 / 2 \log \mathbb{E}_{p(\theta, s)}\left[(\theta-f(s))^2\right] \\
&-1 / 2 \log (2 \pi e),
\end{aligned}
\label{eq:ba-bound-regress}
\end{equation}
where a smaller mean square error produces a larger MI lower bound. This implies that summary statistics that can more accurately predict the parameters in a regression task may contain more information about those parameters. Note that if we further assume the estimator is unbiased and efficient, we can also derive a similar relation between MI and Fisher information \citep{mi-fisher}. In our experiments, we train Light Gradient Boosting Machines (LGBM; \citealp{lgbm}) to predict parameters directly from data summaries. We use the regression performance to validate the MI estimation from MAFs.

\section{Data}
\label{sec:data}
% \subsection{Mock 21 cm SKA images}
 As a demonstration of the this method, we consider an inference problem in 21 cm cosmology, where we need to constrain two Reionization parameters based on mock SKA (Square Kilometre Array) images.  The 21 cm lightcones are simulated using the publicly available code {\tt 21cmFAST}\footnote{https://github.com/andreimesinger/21cmFAST} \citep{Mesinger2007,Mesinger2011}. The simulations were performed on a cubic box of 100 comoving Mpc on each side, with $66^3$ grid cells. In this work, we use coeval boxes at redshift 11.76. We consider the following reionization parameters in simulations and inferences:\begin{enumerate*}

 \item[(1)] $\zeta$, the ionizing efficiency. We vary $\zeta$ as $10 \le \zeta \le 250 $.

\item[(2)] $T_ { \mathrm { vir } }$, the minimum virial temperature of halos that host ionizing sources. We vary this parameter as $ 4 \le \log _ { 10 } \left( T_{ \mathrm { vir } } / \mathrm { K } \right) \le 6 $.
 \end{enumerate*}

 To include observational effects, signals with three different levels of signal contamination are considered here: \begin{enumerate*}
	\item[(i)] signal with ${\bf k}_\perp = 0$ mode removal (removal of the mean for each frequency slice); 
	\item[(ii)] signal with the SKA thermal noise;
	\item[(iii)] signal with the SKA thermal noise $+$ residual foreground after foreground removal with the singular value decomposition.
\end{enumerate*} The thermal noise is produced using the {\tt Tools21cm}\footnote{https://github.com/sambit-giri/tools21cm} \citep{tools21cm} package by considering the SKA1-Low configuration. The foreground simulation is based on the GSM-building model in \citet{zheng2017improved}. We generate 10000 data cubes for each case with different reionization parameters and randomized observational effects.

In this work, we try to evaluate the effectiveness of three different summary statistics in the inference reionization parameters: power spectrum (PS), bispectrum (BS) and scattering transform (ST). For BS, we find that it may contain many uninformative features, which can cause overfitting issues. To address this problem, we conduct feature selection using LGBM by comparing feature importance for parameter estimation in regression tasks. Note that the MI estimates after feature selection are still lower bounds due to data processing inequality. The details of these summary statistics and the process of feature selection are given in Appendix~\ref{apsec:sum}.

\begin{figure*}%[ht]
% \vskip 0.2in
% \begin{center}
\includegraphics[width=2\columnwidth]{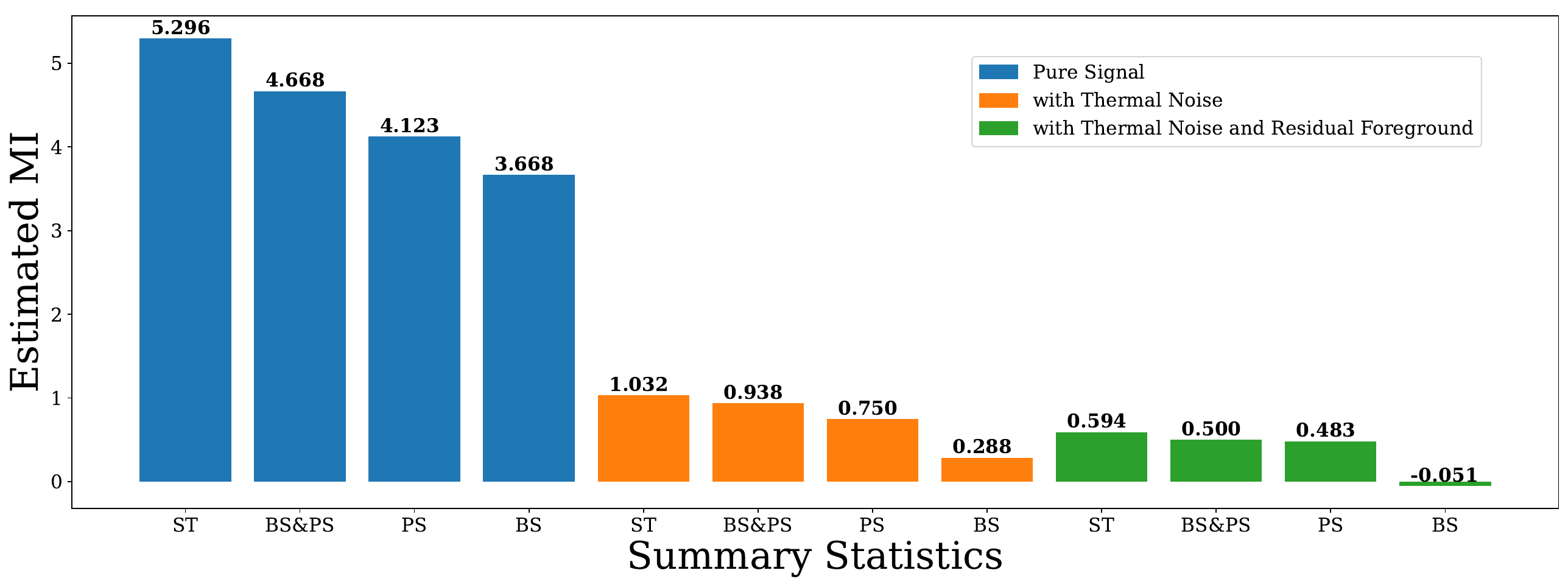}
\caption{The estimated mutual information (MI) between target parameters and various summary statistics under three levels of contamination: (1) pure signal (blue bars) obtained by removing only the ${\bf k}_\perp=0$ mode, (2) signal contaminated by thermal noise from the SKA telescope (orange bars), and (3) signal contaminated by both thermal noise and residual foreground (green bars). We estimated MI for four different statistics for each dataset: power spectrum (PS), bispectrum (BS), scattering transform (ST), and a combination of BS and PS (BS\&PS).}
\label{MI-comp}
% \end{center}
% \vskip -0.2in
\end{figure*}

\section{Results}
\textbf{Comparisons between different summary statistics:} We present our MI estimation results for different data summaries and contamination levels in Figure~\ref{MI-comp}. Our results indicate that the presence of observational effects can significantly reduce the amount of information available about reionization parameters in our mock images. As we increase the level of contamination, the estimated MI decreases accordingly. Furthermore, we find that in all three datasets used in our experiments, the ST is better at extracting physical information than correlation functions. This is an expected result since ST is designed to capture more spatial information \citep{cheng2020new}. The performance of the correlation functions is consistent with previous studies \citep{Watkinson_bs_eor_infer,Shimabukuro_bs_eor_infer_fisher}, where PS and BS are evaluated in a similar inference task. We also notice that the MI estimation of BS in the highest contamination case is negative, while MI should be non-negative by definition. This is because we are estimating a lower bound and the variational distribution is not exact. This result indicates that BS in this case provides nearly no information about the reionization parameters.

% We first show our MI estimation results for different data summaries and different level of data contamination in Figure~\ref{MI-comp}. The fist thing we can see from the results is that observational effects can greatly reduce the amount of information of reionization parameters in our mock images. As expected, the estimated MI reduces as we add stronger data contamination. Secondly, in all three dataset we use in our experiments, we find ST can better extract physical information than correlation functions. This is not surprising since ST is expected to capture more spatial information \citep{cheng2020new}. The order of correlation functions is consistent with our theoretical interpretation, which indicates our method can properly estimate the effectiveness of these summaries statistics. The results seem to also agree with previous study from \cite{Watkinson_bs_eor_infer}, where the authors evaluate the performance of different correlation functions in a similar inference task. 

\begin{figure}%[ht]
% \vskip 0.2in
% \begin{center}
\includegraphics[width=\columnwidth]{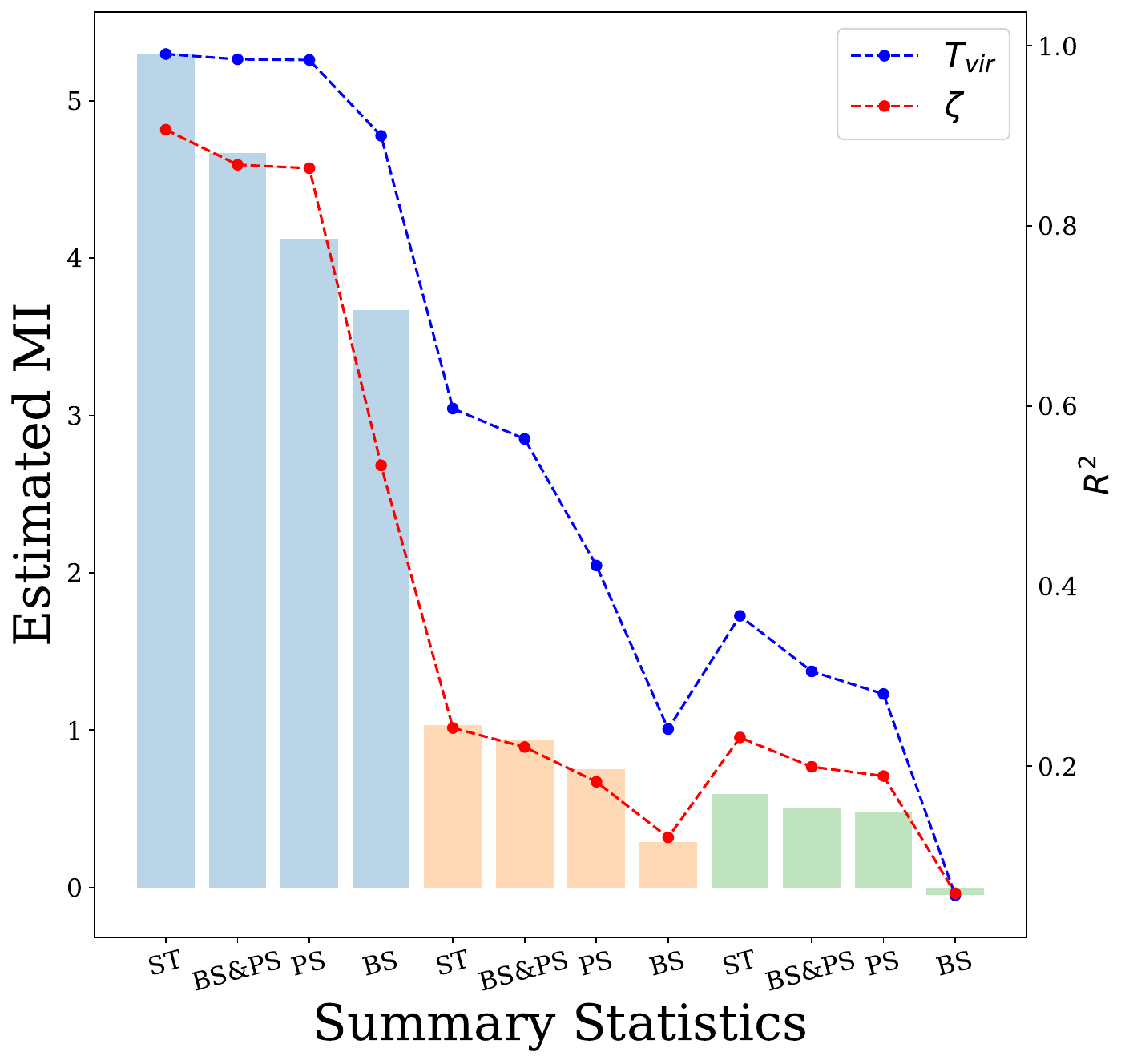}
\caption{Comparison between the optimal $R^2$ achieved by different statistics and their MI estimates. The $R^2$ values(right y-axis) are presented as points overlaid on the MI results. The regression performance for two target parameters, $T_{vir}$ and $\zeta$, is indicated by the blue and red color, respectively.}
\label{Val_R2}
% \end{center}
% \vskip -0.2in
\end{figure}

\textbf{Validation with regression tasks:} Our results are largely consistent with theoretical interpretations and previous works. To further validate their correctness, we trained LGBM to predict reionization parameters and evaluated the regression performance for each dataset. As mentioned in Section~\ref{subsec:unified_MI}, regression performance can also be used as an estimator for mutual information by selecting a specific variational distribution. We used the $R^2$ score to evaluate the regression performance of LGBM and present the results in Figure~\ref{Val_R2}, where we also plotted the previous MI estimates. The two MI estimators are observed to be consistent with each other. However, we note that the $R^2$ score was unable to show the difference between summary statistics when mutual information was relatively high. 

% Our results are generally consistent with theoretical interpretations and previous works. To further verify their correctness, we train LGBM to prediction reionization parameters and evaluate the regression performance for each dataset. As shown in Section ~\ref{subsec:unified_MI}, regression performances can also be treated as a estimator for mutual information by choosing a certain form of variational distribution. We use $R^2$ to evaluate the regression performance of LGBM and show the results in Figure~\ref{Val_R2}, where we also plot the previous MI estimations. We can see that these two lower bounds are consistent with each other. However, $R^2$ fails to show the difference between summary statistics when mutual information is relatively high.

% \textbf{Effect of feature selection:} As discussed in Section~\ref{sec:data}, we use LGBM importance to select features from BS and BS\&PS in all experiments in order to mitigate the impact of overfitting. The MI estimations obtained without feature selection for BS (BS\&PS) for three levels of contamination are 2.090 (4.344), 0.343 (0.938), and -0.006 (0.443), respectively. Compared to the results shown in Figure~\ref{MI-comp} (3.668 (4.668), 0.288 (0.938), -0.051 (0.500)), we observe that the MI estimations after feature selection are generally higher. However, we also notice that for some cases, the results slightly decrease, likely due to the loss of information after feature selection.

\section{Summary}
Our study presents a practical framework for evaluating the effectiveness of summary statistics in a specific inference problem. We accomplish this by estimating the mutual information between these statistics and the target parameters. Unlike existing approaches that rely on inference performance assuming fiducial parameters, our method provides a more robust assessment. We validate this methodology by applying it to the task of inferring reionization parameters from simulated SKA images. Our results demonstrate that the MI estimates agree with previous works and regression-based verification. This novel framework introduces a valuable tool for evaluating the informativeness of summary statistics in the field of cosmology.

% \textbf{Future directions:} Our results indicate that overfitting may bias the MI estimation, and we mitigate this issue by leveraging LGBM's feature selection capabilities. However, we observe that this approach may lead to a loss of information in some cases, implying the need for a more sophisticated feature selection approach. Additionally, given that we estimate only the lower bounds of MI, we emphasize the need for a better verification method to improve our conclusions.

% \begin{figure}%[ht]
% \vskip 0.2in
% % \begin{center}
% \includegraphics[width=\columnwidth]{feature_selection}
% \caption{Effect of feature selection}
% \label{Val_R2}
% % \end{center}
% % \vskip -0.2in
% \end{figure}

% Authors are kindly asked to make their submissions as accessible as possible for everyone including people with disabilities and sensory or neurological differences.
% Tips of how to achieve this and what to pay attention to will be provided on the conference website \url{https://tomworkshop.github.io/}.

% \section*{Software and Data}

% If a paper is accepted, we strongly encourage the publication of software and data with the
% camera-ready version of the paper whenever appropriate. This can be
% done by including a URL in the camera-ready copy. However, \textbf{do not}
% include URLs that reveal your institution or identity in your
% submission for review. Instead, provide an anonymous URL or upload
% the material as ``Supplementary Material'' into the CMT reviewing
% system. Note that reviewers are not required to look at this material
% when writing their review.

\section*{Acknowledgements}

This work is supported by the National SKA Program of China (grant No. 2020SKA0110401), NSFC (grant No. 11821303), and the National Key R\&D Program of China (grant No. 2018YFA0404502). CS thanks Richard Grumitt for useful comments. We acknowledge the Tsinghua Astrophysics High-Performance Computing platform at Tsinghua University for providing computational and data storage resources that have contributed to the research results reported within this paper. 

% \textbf{Do not} include acknowledgements in the initial version of
% the paper submitted for blind review.

% If a paper is accepted, the final camera-ready version can (and
% probably should) include acknowledgements. In this case, please
% place such acknowledgements in an unnumbered section at the
% end of the paper. Typically, this will include thanks to reviewers
% who gave useful comments, to colleagues who contributed to the ideas,
% and to funding agencies and corporate sponsors that provided financial


\bibliography{ref}
\bibliographystyle{tom2023}

%%%%%%%%%%%%%%%%%%%%%%%%%%%%%%%%%%%%%%%%%%%%%%%%%%%%%%%%%%%%%%%%%%%%%%%%%%%%%%%
%%%%%%%%%%%%%%%%%%%%%%%%%%%%%%%%%%%%%%%%%%%%%%%%%%%%%%%%%%%%%%%%%%%%%%%%%%%%%%%
% APPENDIX
%%%%%%%%%%%%%%%%%%%%%%%%%%%%%%%%%%%%%%%%%%%%%%%%%%%%%%%%%%%%%%%%%%%%%%%%%%%%%%%
%%%%%%%%%%%%%%%%%%%%%%%%%%%%%%%%%%%%%%%%%%%%%%%%%%%%%%%%%%%%%%%%%%%%%%%%%%%%%%%
\newpage
\appendix
\onecolumn
\section{Summary Statistics}
\label{apsec:sum}
In this work, we try to evaluate the effectiveness of three different summary statistics in the inference of reionization parameters: power spectrum, bispectrum and scattering transform. 

Power spectrum (PS) $P_{21}({k})$ is the most commonly used statistic, defined as:
\begin{equation}
    \left\langle\delta_{21}(\boldsymbol{k}) \delta_{21}\left(\boldsymbol{k}^{\prime}\right)\right\rangle=(2 \pi)^3 \delta^D\left(\boldsymbol{k}+\boldsymbol{k}^{\prime}\right) P_{21}(\boldsymbol{k}),
\end{equation}
where $\delta_{21}({k})$ is the Fourier transform of the 21 cm brightness temperature field, $\delta^D$ is the Dirac delta function and $\langle...\rangle$ represents an ensemble average. We use the {\tt Tools21cm} \citep{tools21cm} package to calculate the spherical averaged power spectrum of the mock observation.

 Bispectrum (BS) is the Fourier dual of three-point correlation function, defined as:
\begin{equation}
\begin{aligned}
    \left\langle\delta_{21}\left(\boldsymbol{k}_1\right) \delta_{21}\left(\boldsymbol{k}_2\right) \delta_{21}\left(\boldsymbol{k}_3\right)\right\rangle
    =(2 \pi)^3 \delta\left(\boldsymbol{k}_1+\boldsymbol{k}_2+\boldsymbol{k}_3\right) B\left(\boldsymbol{k}_1, \boldsymbol{k}_2, \boldsymbol{k}_3\right).
\end{aligned}
\end{equation}
The BS is a function of three k vectors that form a closed triangle ($\boldsymbol{k}_1+\boldsymbol{k}_2+\boldsymbol{k}_3$=0). In this work, we consider only isosceles triangles and use the python package {\tt Pylians}\footnote{https://github.com/franciscovillaescusa/Pylians3}\citep{Pylians} to calculate the bispectrum of mock observation. We normalize BS following \citet{Watkinson_bs_eor_infer} to separate non-Gaussian information. In our experiments we also consider a combination PS and BS by directly concatenating them and refer it as BS\&PS. For BS, we observe that it may contain some uninformative features (here each feature represents the bispectrum value calculated for a specific triangle configuration), which can lead to overfitting issues. To tackle this problem, we employ feature selection using the LGBM. We first add random noise features to the bispectrum and conduct regression with LGBM. The regression process is repeated multiple times to calculate the mean and variance of the importance of the random noise features. Subsequently, we identify all features that fall within the 3-sigma range of the importance of the random features and exclude them from the fitting of MAFs.

% For BS, we find that it may contain many uninformative features(here each feature refers to the bispectrum value calculated for a certain triangle configuration), which can cause overfitting issues. To address this problem, we conduct feature selection using LGBM. To do so, we first add random noise features to the statistics and run regression with LGBM. After convergence, we record the gain of random noise features and statistics features. We run the regression for multiple times and calculate the mean and variance of importance of random features. Then, we marked all features that have importance within 3-sigma region of random feature importance and exclude them in the fitting of MAFs.

The solid harmonic wavelet scattering transform (ST), first introduced by \citet{eickenberg2017solid,eickenberg2018solid}, is a method for compressing data for inference by convolving the original fields with a cascade of solid harmonic wavelets, performing non-linear modulus on the convolved fields, and integrating over all coordinates. It is an effective way to capture information at different scales and orientations, producing coefficients that are invariant to both translation and rotation. The ST is implemented with the {\tt Kymatio} \footnote{https://www.kymat.io/} \citep{2018arXiv181211214A} package.
%%%%%%%%%%%%%%%%%%%%%%%%%%%%%%%%%%%%%%%%%%%%%%%%%%%%%%%%%%%%%%%%%%%%%%%%%%%%%%%
%%%%%%%%%%%%%%%%%%%%%%%%%%%%%%%%%%%%%%%%%%%%%%%%%%%%%%%%%%%%%%%%%%%%%%%%%%%%%%%

\end{document}